# Thermoelectric properties of new Bi-chalcogenide layered compounds

Yoshikazu Mizuguchi, Atsuhiro Nishida, Atsushi Omachi, and Osuke Miura

Tokyo Metropolitan University, 1-1, Minami-osawa, Hachioji 192-0397, Japan.

Abstract

The layered Bi-chalcogenide compounds have been drawing much attention as a new layered superconductor family since 2012. Due to the rich variation of crystal structure and constituent elements, the development of new physics and chemistry of the layered Bi-chalcogenide family and its applications as functional materials have been expected. Recently, it was revealed that the layered Bi chalcogenides can show a relatively high thermoelectric performance ($ZT$ = 0.36 in LaOBiSSe at ~650 K). Here, we show the crystal structure variation of the Bi-chalcogenide family and their thermoelectric properties. Finally, the possible strategies for enhancing the thermoelectric performance are discussed on the basis of the experimental and the theoretical facts reviewed here.

1. Introduction

Thermoelectric energy conversion is a promising technology for solving energy problems because the energy of the waste heat can be directly converted to the electrical energy. So far, the available thermoelectric devises have been limited due to the absence of high performance thermoelectric materials. To estimate the performance of thermoelectric materials, dimensionless figure-of-merit ($ZT$) is generally used. The $ZT$ can be calculated as $ZT = S^2T/\rho\kappa$, where $S$, $T$, $\rho$, and $\kappa$ are Seebeck coefficient, absolute temperature, electrical resistivity, and thermal conductivity, respectively. Hence, a large absolute value of the Seebeck coefficient, low electrical resistivity, and low thermal conductivity are required for a high $ZT$. Typically, in a conventional semiconductor, $S$ is proportional to $T/n$, where $n$ is carrier concentration with a single band approximation. The $\rho$ is proportional to $1/en\mu$, where $\mu$ is carrier mobility. Therefore, insulators (or semiconductors with a large band gap) exhibit large absolute $S$, but their $\rho$ is usually large. In contrast, metallic compounds exhibit low $\rho$, but absolute $S$ of metallic compounds is generally small. Due to this trade-off relationship of the electrical factors ($S$ and $\rho$), the enhancement of $S^2/\rho$ (power factor: $PF$) is quite difficult. The material, which satisfies both large $S$ and low $\rho$, is semiconductors with a narrow band gap [1]. For example, the $Bi_2Te_3$ family possesses such a band structure and shows a high performance of $ZT \sim 1$ at around 300 K due to the large $\mu$ and the suitable $n$ [2,3]. Hence, $Bi_2Te_3$-based compounds have been used as a practical thermoelectric material for a long time [2]. Another strategy for enhancing $ZT$ is the use of the quantum-size effect. The $ZT$ of $Bi_2Te_3$-based system can be enhanced up to 2.4 by fabricating the $Bi_2Te_3/Sb_2Te_3$ super-lattice devise [4,5]. In addition, the quantum-size effect can be expected to positively work in enhancing $ZT$ in the compounds with a *layered structure*. One of the examples is the layered Co oxides, such as $NaCoO_2$ and related layered compounds [6-8]. These compounds show anomalously large $S$ owing to the strong electron correlations, and the $\kappa$ of these compounds is reduced by the phonon scattering at the interface of the layers. In addition, layered compounds typically possess the great flexibilities of stacking structure and constituent elements as demonstrated in the Co oxides [6-8], which results in the desirable tuning of the electronic structure and the local crystal structure. Therefore, one can say that the layered structure is greatly suitable for designing high $ZT$ materials.

In 2012, we discovered new layered superconductors whose crystal structure is composed of the alternate stacks of the $BiS_2$ bilayer (electrically *conducting layer*) and the electrically insulating *blocking layers* [9-11]. The parent phase (for example, $LaOBiS_2$ with the structure of Fig. 1(a)) is a semiconductor with a band gap [9,12]. When electron carriers were generated in the Bi-6p orbitals by partial substitutions of O by F (in the $La_2O_2$ blocking layers), the F-substituted compounds ($LaO_{1-x}F_xBiS_2$) becomes metallic and shows superconductivity at low temperatures [10,12]. Furthermore, the $La_2O_2$ blocking layer can be replaced by the $RE_2O_2$ layers (RE: rare earth or Bi) or other oxide (or fluoride) layers as shown in Fig. 1. Indeed, the crystal structure can be flexibly modified by changing the blocking layer structure in the BiS$_2$-based compound family [11]. Focusing on the electronic state, we noted that the calculated bad gap were relatively small, and some parent compounds showed low electrical resistivity (as compared normal semiconductors). In addition, the thermal conductivity of the LaO$_{1-x}$F$_x$BiS$_2$ samples were relatively small [13]. Thus, we considered that the BiS$_2$-based layered compounds could be a layered thermoelectric material family as the Bi-Te family or the layered Co oxides family. In this article, the crystal structure variation of the layered Bi-chalcogenide family and the physical properties of those Bi chalcogenides are reviewed. Finally, the possible strategies for enhancing thermoelectric properties of the Bi chalcogenides are discussed.

This article contains new (unpublished) results on CeO$_{1-x}$F$_x$BiS$_2$, NdO$_{1-x}$F$_x$BiS$_2$, and LaOBiPbS$_3$. Therefore, we briefly explain the experimental procedures. The polycrystalline samples were prepared using a conventional solid-state-reaction method with reaction temperatures of 700-800 ºC. The purity and the crystal structure of the obtained samples were investigated using powder X-ray diffraction. The reaction or the annealing processes were carried out in an evacuated quartz tube. Temperature dependence of $\rho$ and $S$ were measured using a four-terminal method with ZEM-3 (Advance Riko).

2. Crystal structure of layered Bi chalcogenides

Typical crystal structures of the layered Bi chalcogenides are summarized in Fig. 1. All the compounds have a layered structure composed of the alternate stacks of the electrically conducting layer and the electrically insulating blocking layer. The typical structures are categorized into the tetragonal (*P*4/*nmm* or *I*4/*mmm*) space group. These Bi chalcogenides have NaCl-type Bi-Ch conducting layers: Bi$_2$Ch$_4$ layer (BiCh$_2$ bilayer) or M$_4$Ch$_6$ layer (M: Bi, Pb). Since the conducting layers contain a two-dimensional Bi-Ch square lattice, the layered Bi chalcogenides exhibit two-dimensional electrical transport.

Figures 1(a-c) are the crystal structures of BiS$_2$-based compounds with the BiS$_2$ bilayer (Bi$_2$S$_4$ layer) as a conducting layer. Among them, the REOBiCh$_2$ (RE: rare earth or Bi; Ch: S or Se) structure (Fig. 1(a)) is the most popular one. In this structure, the Bi$_2$S$_4$ layer is a conducting layer, and the RE$_2$O$_2$ layer acts as a blocking layer. When electron carriers were doped by partial substitutions of O by F, the RE(O,F)BiCh$_2$ compounds become a superconductor with a superconducting transition temperature ($T_c$) as high as 11 K [10,11]. The RE site of RE(O,F)BiS$_2$ can be occupied by La, Ce, Pr, Nd, Sm, Yb, and Bi [10,14-22]. In the case of RE = La, the RE site can be partially substituted with Y, Ti, Zr, Hf, and Th [23,24]. In addition, the S site can be substituted by Se. The RE(O,F)BiS$_{2-x}$Se$_x$-type compounds were reported for RE = La, Ce, and Nd [25-29]. Furthermore, the end member La(O,F)BiSe$_2$ can be synthesized for RE = La [30].

The $RE_2O_2$ layer of the $REOBiS_2$ structure can be replaced with the $Sr_2F_2$ or $Eu_2F_2$ layer, which results in the $SrFBiS_2$ or $EuFBiS_2$ compound [31,32]. The Sr (or Eu) site can be partially substituted by RE; the RE substitution dopes electron carriers into the $BiS_2$ layer [34,35]. In addition, the Sr site can be partially substituted by Ca [32]. Owing to these great flexibility on element substitutions for the RE and Ch sites, the $REOBiCh_2$-type structure has been actively studied, and is particularly important to understand the physical properties of the layered Bi-chalcogenide family.

Figure 1(b) is a crystal structure of $Eu_3F_4Bi_2S_4$ [36] with a $Eu_3F_4$ blocking layer. The $Eu_3F_4$ layer can be regarded as the double $Eu_2F_2$ layers combined to each other with Eu-site sharing. The material itself is a superconductor, and the superconducting $T_c$ was enhanced by Se substitution in $Eu_3F_4Bi_2S_{2-x}Se_x$ [37].

Figure 1(c) is a crystal structure of $Bi_4O_4SO_4Bi_2S_4$, whose structure can be regarded as the alternate stacks of the $Bi_4O_4SO_4$ blocking layer and the $Bi_2S_4$ conducting layer ($BiS_2$ bilayer). It has been considered that the $SO_4$ site in the blocking layer can have some defects, which provides electron carriers into the $BiS_2$ layers [9]. Indeed, the $Bi_4O_4S_3$ compound with 50% $SO_4$ defects (namely, $Bi_4O_4(SO_4)_{0.5}Bi_2S_4$) becomes a superconductor with $T_c \sim 5$ K [9]. Due to the difficulty in determining the $SO_4$ site structure using polycrystalline samples and the absence of single crystals, the precise determination of the $SO_4$ site structure (and the composition) has not been achieved yet [38]; some reports suggested the $Bi_3O_2S_3$ (= $Bi_4O_4S_2Bi_2S_4$) phase was also formed and showed superconductivity [39,40].

Recently, a new Bi-chalcogenide compound with a thick conducting layer of $M_4S_6$ (M = Bi, Pb) was synthesized. Figure 1(d) shows the crystal structure of $LaOBiPbS_3$ [41]. In this structure, an NaCl-type M-S block (layer) is intercalated at between $BiS_2$ layers. Also, this conducting layer can be regarded as the 4-layer-type structure. Thus, the crystal structure variation can be developed by fabricating the multi-layer-type conducting layers as well as by changing the blocking layer structure as demonstrated in the structures of Fig. 1(a-c). We note that the distorted-NaCl-type $M_4S_6$ conducting layer is structurally similar to the $Bi_4Te_6$ conducting layer of $CsBi_4Te_6$, which is a known thermoelectric material (Fig. 1(e)) [3], although the space group of $CsBi_4Te_6$ (monoclinic $C2/m$) is different from that of $LaOBiPbS_3$ and the $BiS_2$-based compounds (tetragonal $P4/nmm$ or $I4/mmm$). The structural symmetry lowering from tetragonal to monoclinic was recently revealed in a single crystal of $LaOBiS_2$ by synchrotron X-ray experiments [42]. Although the X-ray diffraction (XRD) with the polycrystalline $LaOBiS_2$ samples suggested the space group of tetragonal $P4/nmm$, the crystal structure of the $LaOBiS_2$ single crystal was determined to be monoclinic $P2_1/m$. This different results in between polycrystalline and single crystal samples would indicate the crystal structure instability in the $BiS_2$-based compounds. At the same time, the structure instability can be regarded as the *structure flexibility* in the same group of compounds. Thus, we consider that the Bi-chalcogenide family possesses great flexibility of crystal structure (including space group, stacking sequence, and constituent elements), and hence, it is very useful to explore new materials with a high thermoelectric performance.

3. Thermoelectric properties of LaOBiS$_2$-based compounds

3-1. LaO$_{1-x}$F$_x$BiS$_2$

Here, we focus on the LaOBiS$_2$ systems. As introduced above, LaOBiS$_2$ is one of the parent phases of the BiS$_2$-based superconductor, and shows semiconducting-like electrical transport below the room temperature [10]. Band calculations suggested that the LaOBiS$_2$ is a semiconductor with a relatively narrow band gap (< 1 eV) [12]. Partial substitutions of the O site by F generate electron carriers in the BiS$_2$ conducting layers, and the F-substituted LaO$_{1-x}$F$_x$BiS$_2$ shows a superconducting transition at low temperatures [10]. We considered the LaOBiS$_2$-based compounds could exhibit a high thermoelectric properties since we observed a low thermal conductivity in the LaO$_{1-x}$F$_x$BiS$_2$ samples ($\kappa \sim$ 2 W/mK at 300 K) [13]. Thus, we measured high-temperature thermoelectric properties ($\rho$, $S$, and $PF$) of LaO$_{1-x}$F$_x$BiS$_2$ and investigated the effect of the F substitution (electron doping) to the thermoelectric properties [43].

Figures 2(a-c) show the temperature dependences of (a) $\rho$, (b) $S$, and (c) $PF$ for LaO$_{1-x}$F$_x$BiS$_2$. As depicted in Fig. 1(d), the electron carrier concentration is expected to increase with increasing F concentration. The $\rho$ of $x = 0$ increases with increasing temperature, and an anomaly (a hump) is observed at around 500 K. Although the origin of the anomaly has not been clarified yet, it may be related to the (local) structure distortion because the crystal structure of LaOBiS$_2$ can be distorted into monoclinic as revealed in the single crystal structure analysis using synchrotron X-ray [42]. With increasing F concentration, the values of $\rho$ decrease at whole temperatures. The $S$ was negative for all the F concentration at whole temperatures (tested in these experiments), which indicates that the mainly-contributing carrier is electron in this system. The absolute value of $S$ increases with increasing temperature. One of the important facts is the absolute $S$ largely decreases with increasing F concentration. As a result, the values of the $PF$ rapidly decrease with F substitution as shown in Fig. 2(c). Indeed, in the LaOBiS$_2$ system, the electron doping obviously degrades the thermoelectric properties [13,43]. To enhance the thermoelectric performance ($PF$) in the LaOBiS$_2$ system, decreasing $\rho$ without degradation of the absolute $S$ is required. Thus, we next investigated the effects of partial substitutions of S by Se in the conducting layers.

3-2. LaOBiS$_{2-x}$Se$_x$

The S site of LaOBiS$_2$ can be partially substituted by Se [44]. In LaOBiS$_{2-x}$Se$_x$, both S and Se have the same valence of -2 (S$^{2-}$ and Se$^{2-}$). Hence, the Se substitution does not affect the valence of Bi and does not dope electrons, but it should affect the band structure because of the difference of ionic radii of $S^{2-}$ (184 pm) and $Se^{2-}$ (198 pm). Upon the substitution of larger $Se^{2-}$, the enhancement of the orbital overlap between Bi and Ch should be expected.

Figures 3(a-c) show the temperature dependences of (a) $\rho$, (b) $S$, and (c) $PF$ for $LaOBiS_{2-x}Se_x$ [44]. Figure 3(d) is the crystal structure of $LaOBiS_{2-x}Se_x$ and the definitions of the Ch1 and the Ch2 sites. With increasing Se concentration, the values of $\rho$ decrease, indicating that the Se substitution enhances electric conductivity in $LaOBiS_{2-x}Se_x$. The values of $S$ are all negative as observed in $LaO_{1-x}F_xBiS_2$, which suggests that the electrons are mainly contributing in electrical transport in $LaOBiS_{2-x}Se_x$. The absolute values of $S$ do not show a remarkable change up to $x = 0.6$, and it slightly decreases at $x = 0.8$ and 1. The small changes in the $S$ with increasing Se concentration imply that the carrier concentration is not largely affected by the Se substitution. In addition, as will be shown later, the densified LaOBiSSe sample shows the $S$ value comparable to that of $LaOBiS_2$. Thus, we consider that the Se substitution does not largely affect carrier concentration, but it enhances metallic conductivity due to the enhanced carrier mobility; actually, the large enhancement of mobility with increasing Se concentration was revealed in our recent Hall measurements [45]. This can be understood with the concept of in-plane chemical pressure effect as demonstrated in $REO_{0.5}F_{0.5}BiCh_2$-type superconductors [46]. The increase of Se concentration in the BiCh plane results in the enhancement of orbital overlaps between Bi and Ch. It can be considered that the enhanced orbital overlap enhances the carrier mobility, and hence, the metallic conductivity is enhanced. Finally, the calculated $PF$ is shown in Fig. 3(c) as a function of temperature. The $PF$ is clearly enhanced with increasing Se concentration, and large $PF$ values exceeding 4 $\mu W/cmK^2$ are observed for $x = 0.8$ and 1 in $LaOBiS_{2-x}Se_x$. Indeed, the Se substitution largely enhances the thermoelectric performance in the $LaOBiS_2$ system.

3-3. Densified LaOBiSSe

The experimental results on $LaOBiS_{2-x}Se_x$ shown above were based on the polycrystalline pellet samples with a typical relative density of 85-90%. Thus, we densified the $LaOBiS_{2-x}Se_x$ samples using a hot-press (HP) instrument under an applied pressure of 50 MPa and annealing temperature of 700 or 800 °C, with which high density samples were obtained (relative density > 97%) [45,47]. Here, we show the thermoelectric properties of LaOBiSSe because the highest performance was attained in the HP-LaOBiSSe sample among $x = 0-1$.

Due to the uniaxial pressure in the densification process, the obtained sample can possess anisotropic crystal (grain) orientation. So, let us firstly mention the crystal orientation of the HP sample. Figure 4(a) shows the XRD patterns for the powder and pellet samples; for the pellet samples, the scattering vector of X-ray is parallel or perpendicular to the pressing (HP) direction ($P_{//}$ or $P_{\perp}$). The definitions of the $P_{//}$ and $P_{\perp}$ directions are shown in Fig. 4(b). The XRD results propose that the HP process does not affect the phase purity because any impurity phases were not generated.

The peak intensities were different in between $P_{//}$ and $P_\perp$, indicating that the obtained sample was oriented as expected. However, we found that the crystal orientation in the HP sample was very weak: the difference of the peak intensities of the ($h$00) and the (00$l$) peaks for $P_{//}$ and $P_\perp$ are not large. (If the sample was completely oriented, only the ($h$00) or the (00$l$) peaks should be observed as in the case of thin films.)

Figure 4(c) shows the temperature dependences of $\rho$ for two measurement directions (measured with currents of $I // P_{//}$ or $I // P_\perp$). The $\rho$ increases with increasing temperature for both, and the values of $\rho$ for $P_\perp$ are clearly lower than those of $P_{//}$, which is consistent to the fact that the $ab$ plane of the LaOBiSSe grains is relatively oriented along the $P_\perp$ direction. The BiS$_2$-based compounds essentially show two-dimensional electrical conduction [48]. Figure 4(d) shows the temperature dependences of $S$. For both samples, negative $S$ is observed. The absolute values of $S$ for $P_{//}$ are slightly larger than those of $P_\perp$, which is also consistent with the difference in the $\rho$ and the crystal orientation. As a result, the $PF$ values for $P_\perp$ are larger than those for $P_{//}$.

To estimate $ZT$, $\kappa$ was also measured for these samples (Fig. 4(e)). The $\kappa$ decreases with increasing temperature, and the values of $\kappa$ for the $P_\perp$ direction are clearly larger than those for the $P_{//}$ direction. This result indicates that the thermal conductivity along the $c$ axis is obviously smaller than that along the $a$ axis. Figure 4(f) shows the temperature dependences of $ZT$ for both measurement directions. The values of $ZT$ for $P_{//}$ are slightly larger than those for $P_\perp$, and the $ZT$ for these directions at the highest temperature (tested in the study) is almost the same. These results suggest that the thermoelectric performance of LaOBiSSe is not largely affected by the crystal orientation because one direction possesses a good (high) $PF$, and the other direction possesses a good (low) $\kappa$. This characteristic on the insensitivity of $ZT$ to the crystal (grain) orientation may be useful when considering practical application of these BiCh$_2$-based materials.

4. Properties of other layered Bi chalcogenides

Up to here, we focused on the properties of LaOBiS$_2$-based systems. As mentioned with Fig. 1, one of the merits of the Bi-chalcogenide layered compound family is the crystal structure variety. Thus, in this section, we briefly introduce the thermoelectric properties of REO$_{1-x}$F$_x$BiS$_2$ (RE = Ce or Nd) and EuFBiS$_2$ with the REOBiCh$_2$-type structure. In addition, the properties of LaOBiPbS$_3$ with the 4-layer-type compound (Fig. 1(d)) at high temperatures are shown.

4-1. REO$_{1-x}$F$_x$BiS$_2$

Figures 5(a-c) show the temperature dependences of $\rho$, $S$, and $PF$ for CeO$_{1-x}$F$_x$BiS$_2$ with $x$ = 0, 0.25, and 0.5. The values of $\rho$ for $x = 0$ (CeOBiS$_2$) are clearly lower than those of LaOBiS$_2$. It has been reported that the polycrystalline sample of CeOBiS$_2$ shows metallic conductivity at low

temperature [14], which seems to be consistent to the present data. In contrast, the single crystal of CeOBiS$_2$ shows semiconducting behavior at low temperatures [49]. On the basis of these diverse experimental facts in polycrystalline samples and single crystals, the electronic states can be different in between these two sample forms. As a fact, the photoemission experiment suggested that the Ce valence is in the mix-valence state of Ce$^{3+}$ and Ce$^{4+}$ [50], which should provide excess electron carriers to the BiS$_2$ conducting layers. Local structure distortion and/or the effect of the grain size may be affecting the Ce valence and the electronic states of polycrystalline CeOBiS$_2$. The low absolute values of $S$ for CeOBiS$_2$ (Fig. 5(b)) seems to be consistent with the mixed-valence scenario in CeOBiS$_2$ because the absolute value of $S$ of semiconductors generally decreases with increasing carrier concentration. As a result, the $PF$ for CeOBiS$_2$ is lower than that of LaOBiS$_2$ and comparable to electron-doped LaO$_{0.95}$F$_{0.05}$BiS$_2$.

For $x = 0.25$ of CeO$_{1-x}$F$_x$BiS$_2$, the values of $\rho$ are larger than those for $x = 0$. The absolute values of $S$ for $x = 0.25$ are also larger than those of $x = 0$. For $x = 0.5$, the values of $\rho$ and absolute $S$ are lower than those for $x = 0$. These experimental results suggest that the effects of F substitutions to the thermoelectric properties in CeO$_{1-x}$F$_x$BiS$_2$ cannot be simply understood. First, we have to consider the effect of the mixed-valence states of Ce. In addition, we are tuning electron carriers by the F concentration. Furthermore, the carrier mobility should largely affect the electrical transport as shown in LaOBiS$_{2-x}$Se$_x$. Actually, the $a$-axis is expanded with increasing F concentration in these CeO$_{1-x}$F$_x$BiS$_2$ samples and in the previous reports as well [14]. By simply estimating the in-plane chemical pressure effect in CeO$_{1-x}$F$_x$BiS$_2$, the elongation of the $a$-axis should result in the decrease in in-plane chemical pressure (decrease in orbital overlap) and carrier mobility. These multiple factors make the understanding of the thermoelectric properties of CeO$_{1-x}$F$_x$BiS$_2$ quite difficult. Anyway, the replacement of the La$_2$O$_2$ blocking layer to the Ce$_2$O$_2$ layer does not positively work in enhancing $PF$.

Next, let us introduce the properties of NdO$_{1-x}$F$_x$BiS$_2$ with $x = 0.25$ and 0.5; unfortunately, the polycrystalline samples of $x = 0$ cannot be obtained in the system. Figures 5(d-f) show the temperature dependences of $\rho$, $S$, and $PF$ for NdO$_{1-x}$F$_x$BiS$_2$. Both samples show similar temperature dependences of $\rho$ and $S$, but the values of $\rho$ for $x = 0.25$ are lower than those for $x = 0.5$ while the nominal composition of doped F is half in the $x = 0.25$ sample. The almost same values of $S$ would suggest that the effective carrier concentration is almost the same. This assumption is consistent with the fact that the superconducting transition temperature of the NdO$_{1-x}$F$_x$BiS$_2$ system is not sensitive to the F concentration within a wide range of $x = 0.1$-0.7 [16,51,52]. The $a$-axis of NdO$_{1-x}$F$_x$BiS$_2$ also increases with increasing F concentration. The larger $\rho$ values for $x = 0.5$ than those for $x = 0.25$ may be resulting from the decrease in the in-plane chemical pressure with the $a$-axis elongation as well as in CeO$_{1-x}$F$_x$BiS$_2$. As facts, the values of $PF$ for NdO$_{1-x}$F$_x$BiS$_2$ are quite low as compared to LaOBiS$_{2-x}$Se$_x$.

As shown here, the replacement of the La$_2$O$_2$ blocking layer to the other RE$_2$O$_2$ blocking layer does not positively work in enhancing *PF*. In addition, the Se substitution for the S site cannot be demonstrated in the Ce- or Nd-based compounds. The reason may be due to the ionic radius of Se$^{2-}$, which would be too large for the Ce$_2$O$_2$ or Nd$_2$O$_2$ blocking layers. Next, we review the recent studies on EuFBiS$_2$ and LaOBiPbS$_3$.

4-2. EuFBiS$_2$

EuFBiS$_2$ crystalizes in the tetragonal *P*4/*nmm* space group, which can be obtained by replacing the La$_2$O$_2$ layer of LaOBiS$_2$ to the Eu$_2$F$_2$ layer (Fig. 6(c)). This material shows metallic conductivity and shows a superconducting transition at 0.4 K [32]. Goto *et al.* measured thermoelectric properties of EuFBiS$_2$ at high temperatures [53]. Figures 6(a,b) show the temperature dependences of $\rho$ and *S* for EuFBiS$_2$. At *T* = 300-623 K, the $\rho$ showed a slight decrease with increasing temperature but the values were almost constant ($\rho$ = 3.5-4 mΩcm). The *S* at 300 K was -32 μV/K, and the absolute value of *S* increased with increasing temperature and reached -50 μV/K at 623 K. The estimated $\kappa$ at 300 K was about 2 W/mK, which was also low and close to that of LaOBiS$_2$. Indeed, low thermal conductivity seems to be a general characteristic of the layered Bi-chalcogenide family. The calculated *PF* and *ZT* were 0.71 μW/cmK$^2$ and 0.02 at 623 K, respectively. We consider that the low *ZT* (*PF*) in EuFBiS$_2$ can be attributed to the excess electron carriers as in the case of CeOBiS$_2$. Zhai *et al.* proposed that the Eu valence in EuFBiS$_2$ is not +2 but +2.1-2.3, and the metallic characteristics were resulting from the excess electron carriers doped in the BiS$_2$ layers due to the mixed-valence state of Eu [32]. The carrier concentration (*n*) estimated from the Hall measurement is 3.2×10$^{21}$ cm$^{-3}$, which is clearly larger than those of the parent phases of REOBiS$_2$. Therefore, a high *ZT* cannot be obtained in EuFBiS$_2$ without compensation of the excess electron carriers in the BiS$_2$ layers.

4-3. LaOBiPbS$_3$

As introduced in Fig. 1, LaOBiPbS$_3$ has the 4-layer-type conducting layer, which is similar to the Bi$_4$Te$_6$ conducting layer of CsBi$_4$Te$_6$ [3,41]. Hence, this material is very important to discuss how we can enhance the thermoelectric properties of the Bi-chalcogenide layered compound family. Sun *et al.* reported thermoelectric properties of LaOBiPbS$_3$ below the room temperature [41]. The $\rho$ at 300 K was ~8 mΩcm, and the $\rho$ largely increased with decreasing temperature (particularly below 100 K). The *S* at 300 K was -50 μV/K. The $\kappa$ at 300 K was ~4 W/mK. Using these data, the *PF* and *ZT* at 300 K can be calculated to be 0.31 μW/cmK$^2$ and 0.0023, respectively.

We examined the thermoelectric properties of LaOBiPbS$_3$ at high temperatures. Figures

7(a-c) show the temperature dependences of $\rho$, $S$ and $PF$ for LaOBiPbS$_3$. The values of $\rho$ are 7-8 mΩcm at between 300 and 740 K, and the $\rho$ does not show a large change with increasing temperature. In contrast, the absolute values of $S$ shows a large increase with increasing temperature. The $S$ at 740 K is -92 μV/K. Hence, the calculated $PF$ largely increases with increasing temperature and reaches 1 μW/cmK$^2$ at 740 K. Although the values of $PF$ for LaOBiPbS$_3$ are still lower than those for LaOBiSSe, the $PF$ will be enhanced by increasing carrier mobility as demonstrated in LaOBiS$_{2-x}$Se$_x$.

5. Summary and possible strategies for a high thermoelectric performance

We reviewed the crystal structure variation and the physical properties of the new Bi-chalcogenide layered compound family. Here, we briefly summarize the evolution of thermoelectric properties by the element substitution or the manipulation of the layered structure and would like to discuss the possible strategies for a high thermoelectric performance in the layered Bi-chalcogenide compound family. One of the advantages of the layered Bi-chalcogenide compounds as a thermoelectric material is the low thermal conductivity. The values of $\kappa$ are quite lower than that of the other chalcogenides or other inorganic materials with electrical conduction. In addition, as shown in the part of LaOBiSSe, the $\kappa$ decreases with introducing disorder as the Se substitution for the S site in LaOBiS$_{2-x}$Se$_x$. Therefore, the thermal conductivity will be tunable at the final stage of the material design of Bi-chalcogenide thermoelectric materials. Thus, we need to enhance $PF$ at the early stage, and now, we are exploring new materials with $PF$ higher than that of LaOBiSSe. On the basis of the results of the electron doping in LaO$_{1-x}$F$_x$BiS$_2$, the excess electron carriers should degrade the $PF$. In addition, the compounds containing self-doped carriers, such as CeOBiS$_2$ or EuFBiS$_2$, shows a lower performance due to the excess carriers as well. Namely, the parent compound with a band gap should be preferable for a high $PF$ material, which seems to be consistent with the Mott relationship ($S$ can be expressed as a function of $T/n$) as used in the study of LaO$_{1-x}$F$_x$BiS$_2$ [13]. Therefore, the lower $n$ would result in a high absolute value of $S$ in this system. To discuss this assumption, we plotted the $n$ and the $S$ reported in Refs. 13, 32, 41, 45, and 53 in Fig. 8. The values of $n$ for these compounds were estimated from the Hall measurements by assuming the single-band model for contributing carriers, which are electrons in the present system. Since the Hall coefficient of the electron-doped compounds could not be explained from the single-band model [13], we excluded those electron-doped compounds from this plot. At least, the data points in Fig. 8 ride on a single slope: the absolute values of $S$ are related to the $n$. Thus, a lower $n$ should be needed for a larger value of absolute $S$. Next, we need to decrease the $\rho$ without decrease of absolute $S$. The strategy to achieve this situation should be the enhancement of in-plane chemical pressure effect, which can increase the carrier mobility by the enhancement of in-plane orbital overlaps. On the basis of this scenario, the $Bi^{3+}$-$Se^{2-}$ or the $Bi^{3+}$-$Te^{2-}$ bonding will be preferable than the $Bi^{3+}$-$S^{2-}$ bonding. In addition, the $Pb^{2+}$ ion will be useful to enhance the in-plane chemical pressure because the $Pb^{2+}$ possesses a large ionic radius. Furthermore, the crystal structure symmetry lowering from tetragonal to monoclinic (or orthorhombic) may be useful to enhance the carrier mobility. The $LaOBiCh_2$ structure can be distorted from tetragonal (*P*4/*nmm*) to monoclinic (*P*2$_1$/*m*). In the monoclinic structure, Bi-Ch zigzag chains form in the conducting plane, which should strongly affect the electronic states [42,54,55]. With these possible strategies for enhancing *PF* in the layered Bi-chalcogenide compounds, we are going to explore a breakthrough material for the achievement of the novel thermoelectric applications.

Acknowledgements

The authors thank C. H. Lee, H. Nishiate, Y. Goto, Y. Kamihara, M. Matoba, K. Kuroki, H. Usui, T. Hiroi, and J. Kajitani for experimental supports and fruitful discussion for the research projects on the thermoelectric properties of the layered Bi chalcogenides. The studies shown in this article were partly supported by Grant-in-Aid for Scientific Research (25707031 and 15H05886) and the TEET research fund (2012 and 2014).

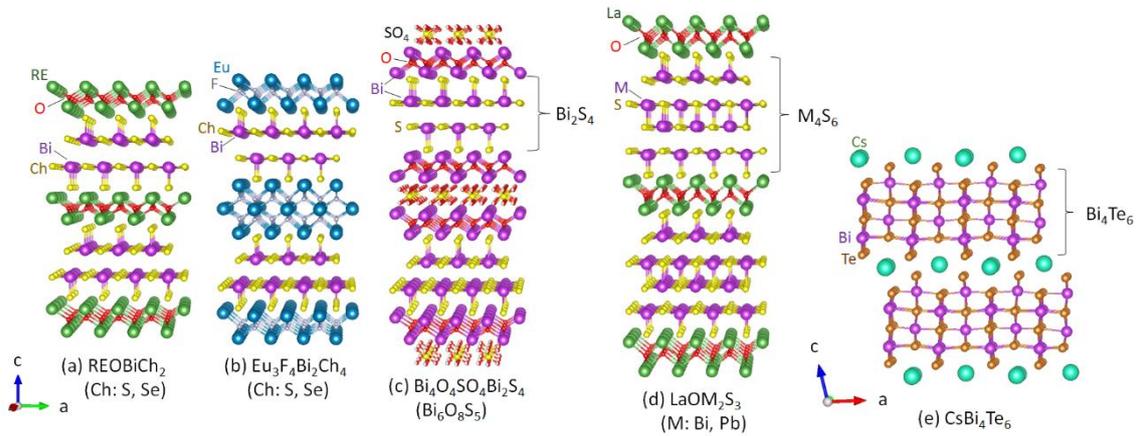

Fig. 1. Schematic images of the crystal structure of various Bi-chalcogenide compounds. (a-c) Typical BiS$_2$-based compounds: (a) REOBiCh$_2$ (RE: rare earth or Bi; Ch: S, Se) [10,30], (b) Eu$_3$F$_4$Bi$_2$Ch$_4$ [36,37], and (c) Bi$_4$O$_4$SO$_4$Bi$_2$S$_4$ [9]. The electrically conducting layer of these compounds is the two-layer-type Bi$_2$S$_4$ layer. (d) LaOM$_2$S$_3$ (M: Bi, Pb) [41]. The M$_4$S$_6$ conducting layer of LaOM$_2$S$_3$ is similar to the Bi$_4$Te$_6$ layer of (e) CsBi$_4$Te$_6$ [3]. The crystal structure images were prepared using VESTA software [56]

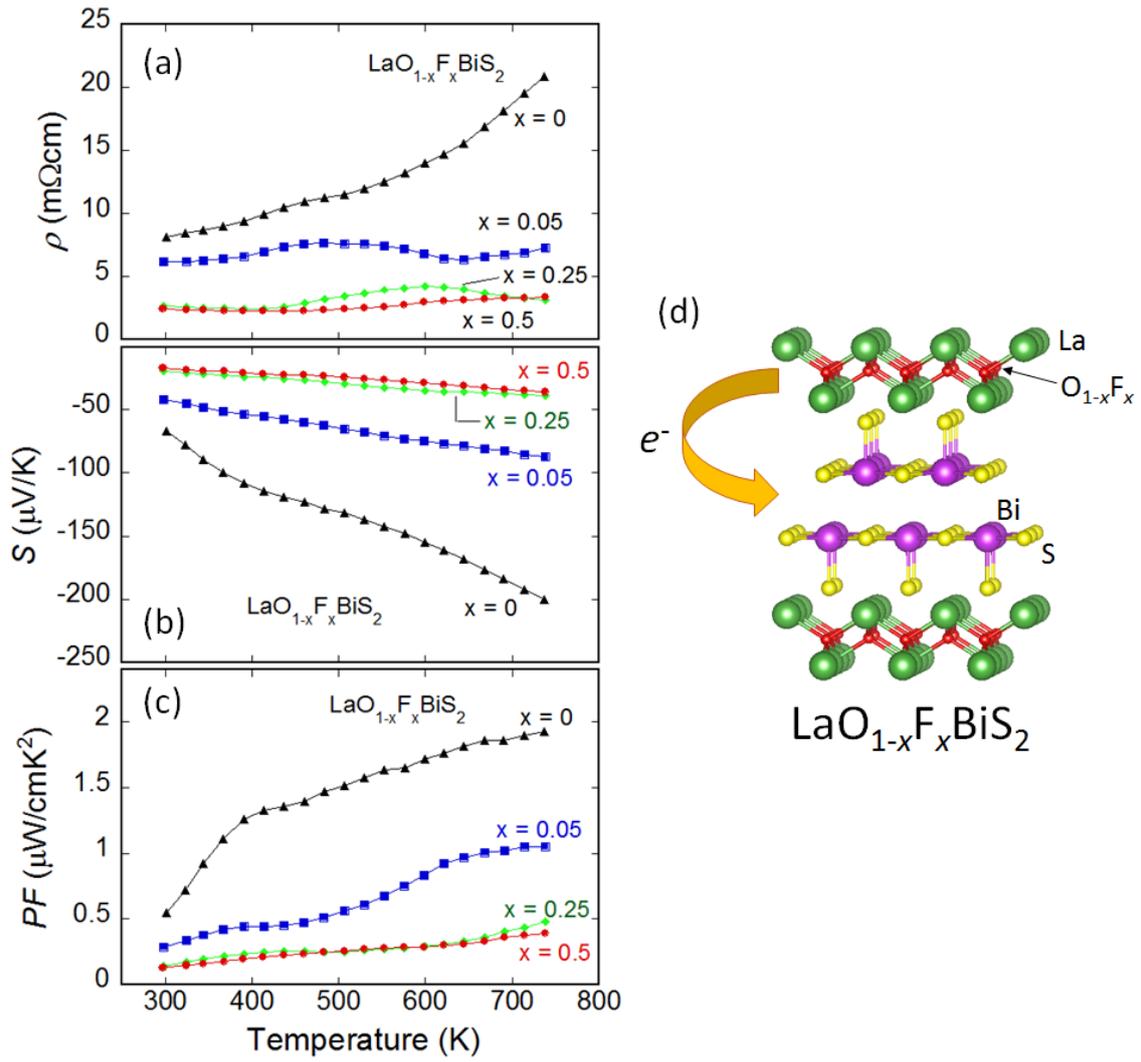

Fig. 2. (a-c) Temperature dependences of (a) electrical resistivity ($\rho$) (b) Seebeck coefficient ($S$), and (c) power factor (*PF*) for LaO$_{1-x}$F$_x$BiS$_2$. (d) Schematic image of the crystal structure of LaO$_{1-x}$F$_x$BiS$_2$ and electron doping scenario. Refer to Ref. 43 (J. Appl. Phys. 115, 083909 (2014)) for original data.

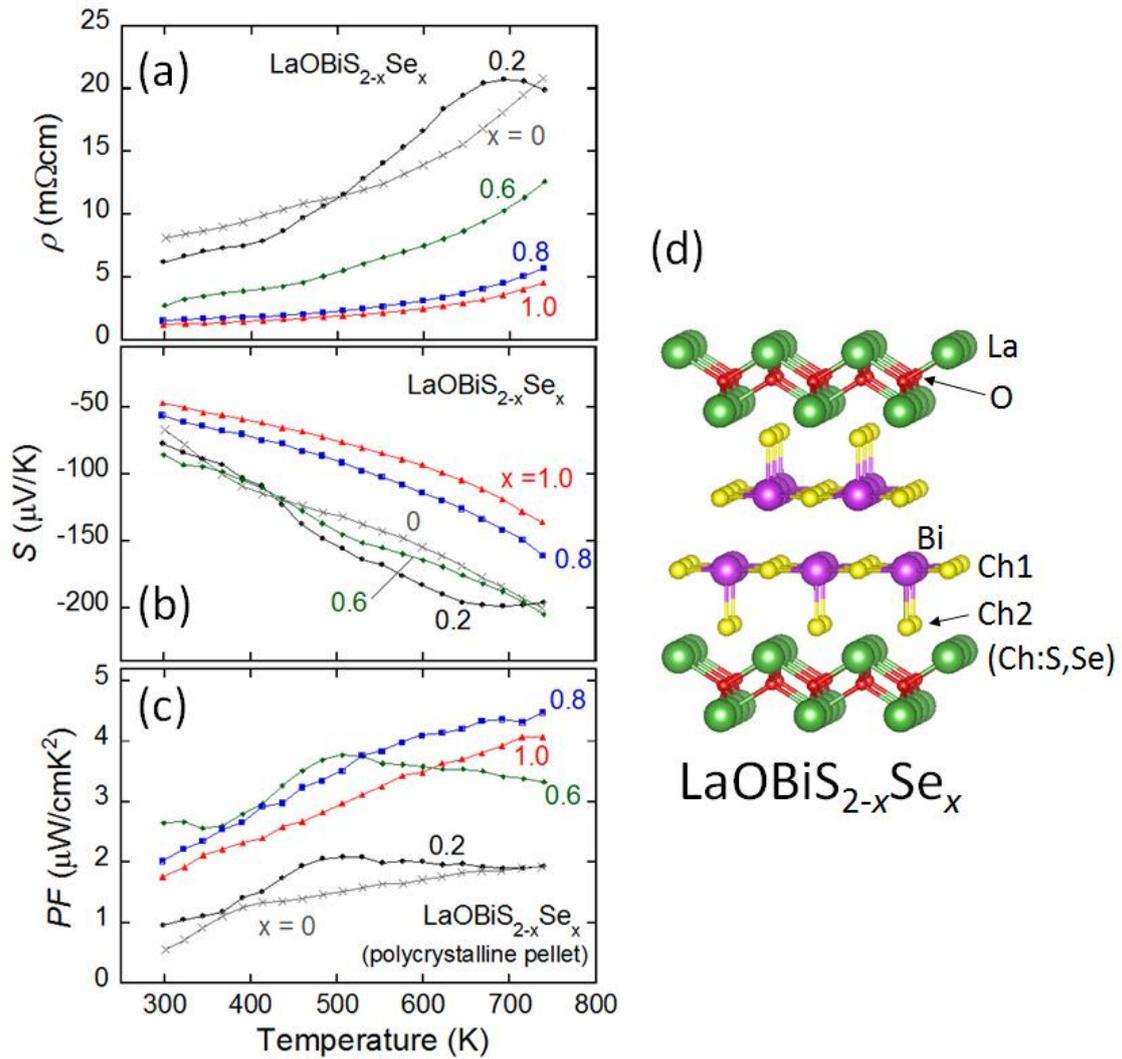

Fig. 3. (a-c) Temperature dependences of (a) electrical resistivity ($\rho$) (b) Seebeck coefficient ($S$), and (c) power factor ($PF$) for LaOBiS$_{2-x}$Se$_x$. (d) Schematic image of the crystal structure of LaOBiS$_{2-x}$Se$_x$ and the definitions of the Ch1 and Ch2 sites. Refer to Ref. 44 (J. Appl. Phys. 116, 163915 (2014)) for original data.

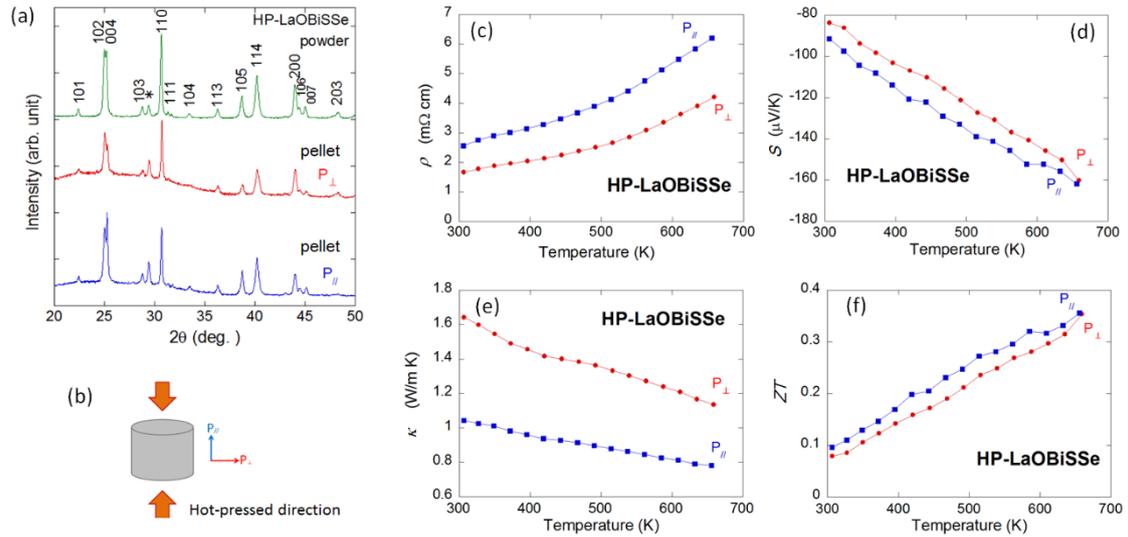

Fig. 4. (a) XRD patterns for hot-pressed (HP) LaOBiSSe ($x = 1$). The Miller indices are shown with the top profile. The asterisk indicates the impurity ($La_2O_3$: 7% against the major phase) peak. To investigate the crystal structure anisotropy, XRD measurements were performed for polished pellets with two scattering vectors of $P_{//}$ and $P_{\perp}$. (b) Schematic image for the definitions of the measurement directions of $P_{//}$ and $P_{\perp}$ and the hot-pressing direction. (c-f) Temperature dependences of (c) electrical resistivity ($\rho$), (d) Seebeck coefficient ($S$), (e) thermal conductivity ($\kappa$), and dimensionless figure-of-merit ($ZT$) for HP-LaOBiSSe. Refer to Ref. 47 (Appl. Phys. Express 8, 111801 (2015)) for original data.

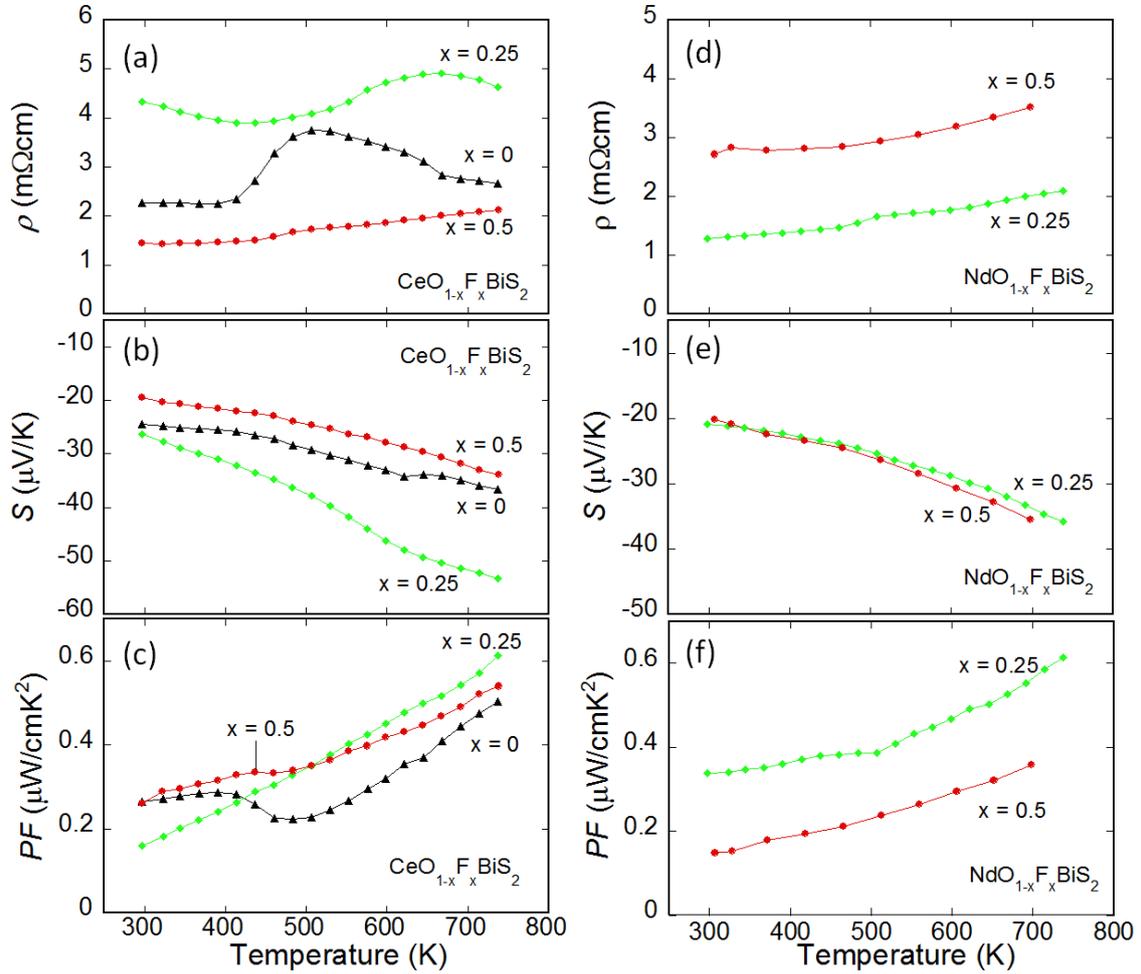

Fig. 5. (a-c) Temperature dependences of (a) electrical resistivity ($\rho$), Seebeck coefficient ($S$), and power factor ($PF$) for $CeO_{1-x}F_xBiS_2$. (d-f) Temperature dependences of (d) $\rho$, (e) $S$, and (f) $PF$ for $NdO_{1-x}F_xBiS_2$. Note that $NdOBiS_2$ ($x = 0$) was not obtained with the solid-state-reaction synthesis.

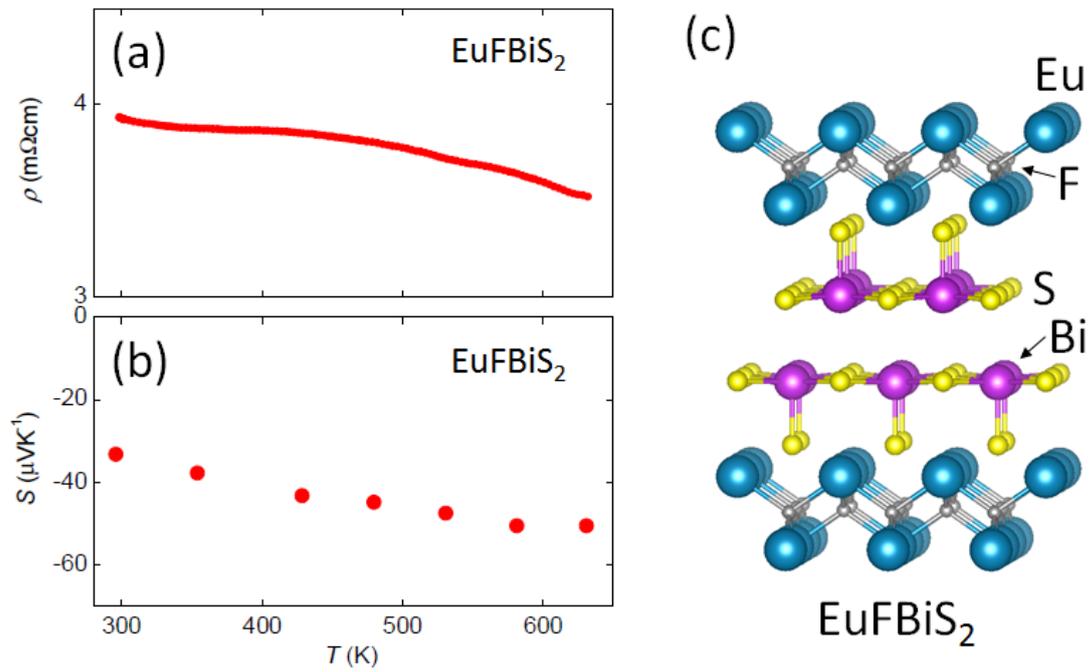

Fig. 6. Temperature dependences of (a) electrical resistivity ($\rho$) and (b) Seebeck coefficient ($S$) for EuFBiS$_2$. (c) Schematic image of the crystal structure of EuFBiS$_2$. Refer to Ref. 53 (J. Phys. Soc. Jpn. 84, 085003 (2015)) for the original data for Figs. 6(a,b).

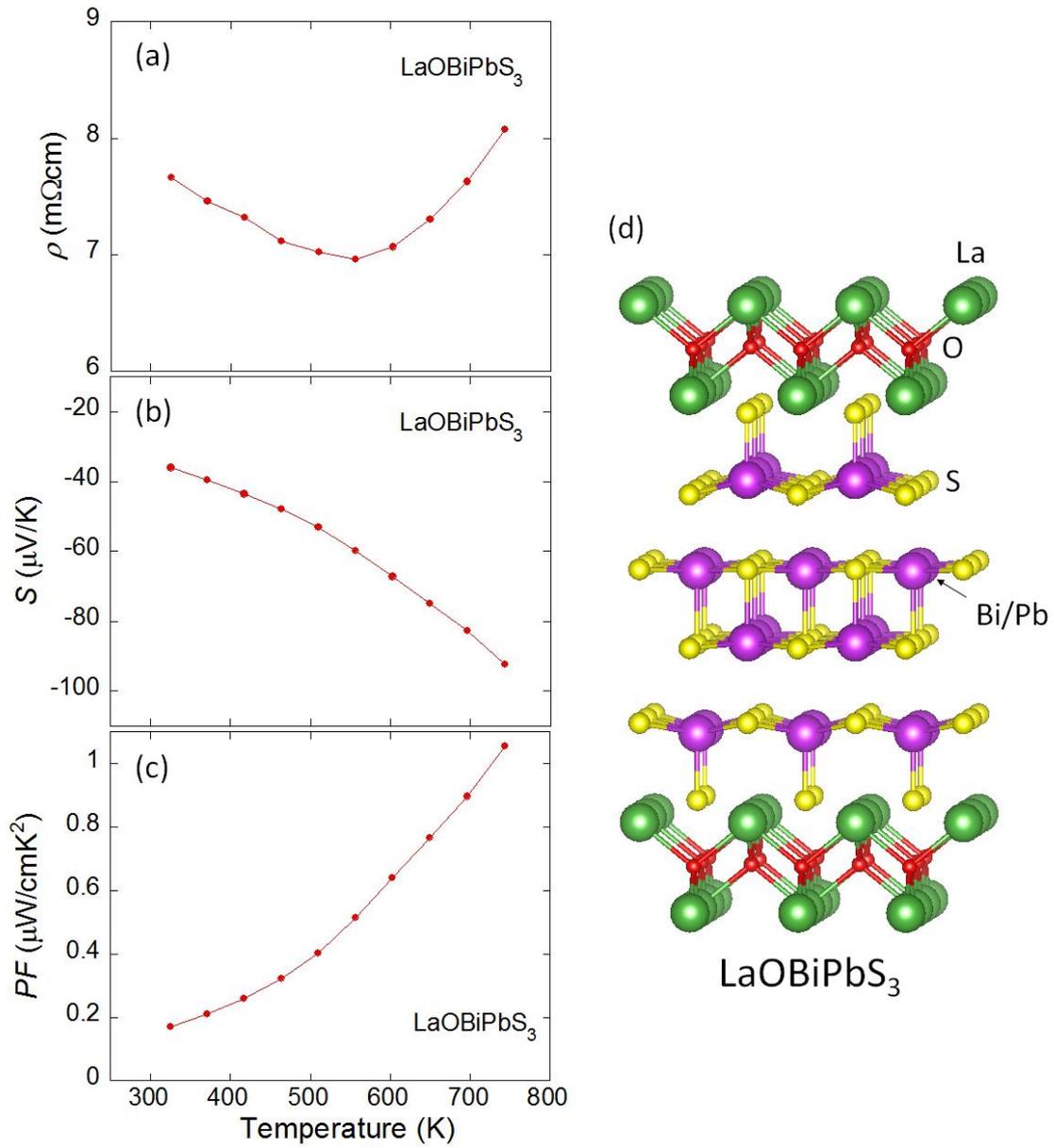

Fig. 7. (a-c) Temperature dependences of (a) electrical resistivity ($\rho$), Seebeck coefficient ($S$), and power factor ($PF$) for LaOBiPbS$_3$. (d) Schematic image of the crystal structure of LaOBiPbS$_3$.

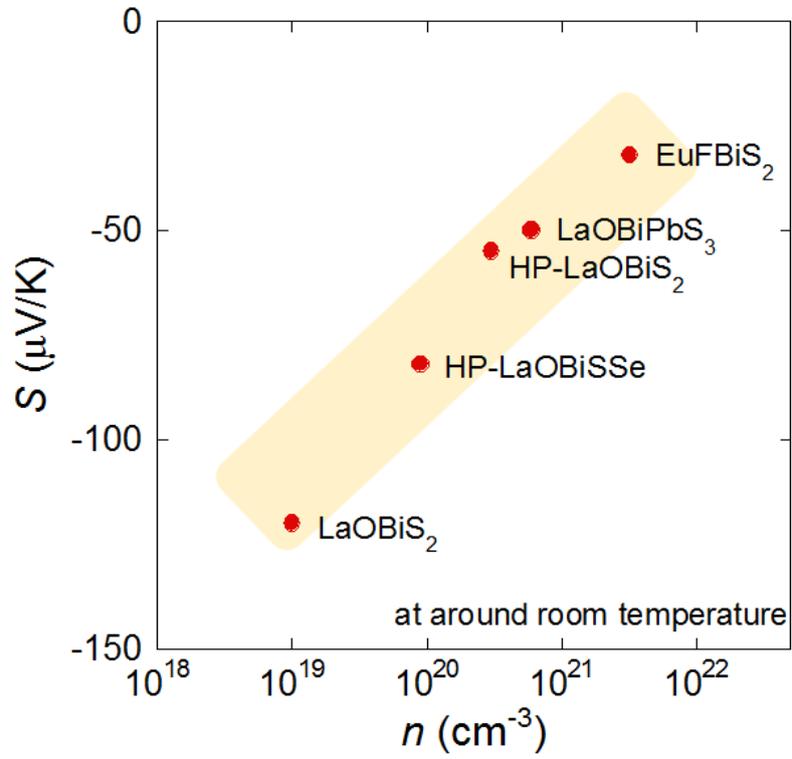

Fig. 8. The Seebeck coefficient (*S*) for several Bi-chalcogenide samples (parent phases) are plotted as a function of the carrier concentration (*n*) (log scale for *n*).